\documentstyle[aps,preprint,prl,epsf]{revtex}
\draft
\begin{document}

\title{Strong Field Ionization in Arbitrary Laser Polarizations}
\author{M. Protopapas$^1$, D.G. Lappas$^{2}$ and P.L. Knight$^1$}
\address{$^1$Optics Section, Blackett Laboratory, Imperial College 
London, London SW7 2BZ, U.K.}
\address{$^2$Department of Physics, Lund Institute of Technology, 
PO Box 118, S-221 00 Lund, Sweden}

\maketitle

\begin{abstract}
We present a numerical method for investigating the non-perturbative 
quantum mechanical interaction of light with atoms in two dimensions, 
without a basis expansion. This enables us to investigate
intense laser-atom interactions with light of arbitrary polarization
without approximation. 
Results are presented for the dependence of ionization and high harmonic 
generation on ellipticity seen in recent experiments. 
Strong evidence of stabilization in circular polarization 
is found.
\end{abstract}

In the last ten years the interaction of intense laser light with atoms
has been of great interest because of the many novel non-perturbative 
phenomena that have been observed, such as Above Threshold Ionization 
(ATI), High Harmonic Generation (HHG) and Stabilization \cite{papas1}. 
The strongly non-perturbative nature of all these phenomena 
makes theoretical analysis difficult. However, considerable success 
has been made in using semi-analytic approaches and numerical methods 
for integrating the Schr\"odinger equation. The numerical approach 
has been used extensively in the past ten years due in part to the 
increase in available computing power, the lack of approximations 
(except for those inherent to the Schr\"odinger equation and the dipole 
approximation) and the rather limited applicability of analytic approaches. 
Despite the success of the numerical methods most investigations have 
dealt with only linearly polarized light: the reason for this is that 
at least two dimensions are needed in the case of arbitrary polarization, 
so established one-dimensional approaches \cite{eberly} are inappropriate. 
Also, current three-dimensional approaches 
\cite{kul,pot,lam,rmat} are unable to deal with the enormous basis 
set needed to represent adequately the evolving wave function unless 
a parallel supercomputer \cite{pir} is used.
The approach used by many authors \cite{kul} to describe atomic
dynamics in strong laser fields expands the atomic wave function
into sums of products of radial wave functions (channels) and 
angular factors. These channels are coupled together by the very 
strong laser field, and for {\it linear} polarization, the channel 
couplings lead to tri-diagonal sparse matrices amenable to efficient
numerical solution. This approach entirely fails for polarizations 
other than linear, as the matrix sparseness properties totally 
disappear \cite{sturm}.

Recent experiments \cite{elli,liang} on the very strong dependence 
of high harmonic generation on polarization ellipticity demand theoretical 
approaches to intense field ionization going beyond the existing methods
which rely on {\em linear} polarizations.
In this letter, we introduce a method for integrating numerically
the two dimensional Schr\"odinger equation without using a restrictive
basis expansion. This enables us to study light of arbitrary polarization 
and realistic frequencies, interacting with an atom. 
Fully quantum mechanical results for ionization and high harmonic 
generation are presented for circular and elliptically polarized light, 
showing the strong ellipticity dependence seen in recent experiments
\cite{elli,liang}. Also, we show that in the intermediate regime 
between tunneling and multiphoton ionization the ionization dependence 
on ellipticity is opposite to that seen with purely tunneling dynamics; 
this may be critical for experiments towards generating attosecond pulses. 

As in all studies of this kind the starting point is the Schr\"odinger
equation (atomic units are used throughout):
\begin{equation}
\mbox{i} \frac{\partial}{\partial t} \Psi(x,y,t) =
\left [ -\frac{1}{2} \left ( \frac{\partial^2}{\partial x^2}
+ \frac{\partial^2}{\partial y^2} \right )
- \frac{1}{\sqrt{a^2 + x^2 + y^2}} 
+ (xE \sin \omega t + \epsilon yE \cos \omega t ) f(t) \right ] 
\Psi(x,y,t), 
\label{sch}
\end{equation}
where $\Psi(x,y,t)$ is the two dimensional wavefunction, $f(t)$ the pulse 
envelope and $\epsilon$ the ellipticity. The second term on the right hand 
side is the smoothed Coulomb potential \cite{eberly} used to avoid numerical 
problems associated with the singularity at $x=y=0$ 
($a=0.8$ so that the ground state has the same binding energy as 
hydrogen, i.e. $-0.5$ a.u.). The pulse envelope, $f(t)$, used here is a 
2-cycle linear turn-on followed by 2 cycles of constant intensity 
and then a 2-cycle linear turn-off, which ensures no residual drift 
motion in the free electron displacement. The ellipticity is defined 
such that $\epsilon=0$ represents linear polarization along the x-axis
and $\epsilon=1$ circular.

Equation \ref{sch} is integrated using a 
method analogous to that implemented by Grobe and Eberly \cite{grobe1} 
and others in their studies of one-dimensional two-electron systems 
in which both electrons are restricted to one dimension. In our case 
there is only one electron with two degrees of freedom. The core of 
the approach is the split step method \cite{feit} in which the time 
evolution operator is split into kinetic and potential operators:
\begin{equation}
U(\Delta t)=\exp[-\mbox{i} T \Delta t/2] \exp[-\mbox{i} V \Delta t]
\exp[-\mbox{i} T \Delta t/2] + O(\Delta t^3),
\end{equation}
where $T$ is the kinetic part of the Hamiltonian (including the 
dipole interaction, in a `minimal coupling' form) and $V$ the 
potential part. The action of the kinetic operator is efficiently 
carried out in Fourier space, while the action of the potential 
operator is carried out in real space. An absorber is used to 
remove any part of the wavepacket reaching 
the boundaries, so that artificial reflections are avoided. It is 
assumed that any wave function reaching the end of the grid box 
represents ionization, and so the remaining norm gives the necessary 
information about the amount of ionization occurring. This means that 
any population in high lying Rydberg states may also be removed; 
however, this is negligible for the parameter regime of interest here, 
as we have verified by varying the box dimensions.
In all the cases presented, the initial wave 
function is the ground state, found by imaginary time 
integration. A more detailed discussion on various computational 
aspects can be found in ref. \cite{grobe1}. 

In Fig. 1 we show the wavepackets generated after 3.5 cycles of 
evolution for the previously described pulse shape with laser light 
of wavelength 526 nm ($\omega=0.0867$ a.u.), i.e., the wavelength of 
a frequency doubled Nd:YAG laser and {\it intensity} $3.51 \times 10^{14} 
Wcm^{-2} (0.01 a.u.)$. For the case of linear polarization it is 
clear that the wavepacket is concentrated about the polarization 
axis, but there is a non-negligible width due to 
transverse wavepacket spreading. Interestingly, side lobes can be 
seen in the wavepacket; these are formed by a combination of the
slow transverse spreading, the relatively fast laser driven motion 
in the polarization direction and the Coulomb attraction (rescattering).

The behaviour of the wavepacket created by circular polarization is 
very different. The packet resembles a spiral with modulations within 
the tail, formed as the wavepacket streams away from the atom down 
the potential barrier in the radial direction. 
As well as being interesting in their own right the wavepackets 
show pictorially that the recollisions necessary for efficient HHG can 
only occur for light with $\epsilon \approx 0$ \cite{elli}. 
The complex structure of the wave function in the linear polarization 
case reflects the fact that rescattering from the atomic core, 
according to the recollision model of HHG \cite{kul2,cork,lew},
produces strong time-dependent interference patterns in the 
vicinity of the atomic core, which are responsible for much of the 
structure in the spectrum of HHG \cite{papas2}. Those interferences 
are absent in the circular polarization case, as is clear from 
Fig. 1(b), due to the fact that the probability of rescattering with 
the atomic core is negligible. We have observed a smooth transition 
from the recollision behaviour resulting from linear polarization to 
the swirling tail of the circular polarization, by integrating the 
wavepacket dynamics for ellipticities ranging from 0 to 1.
This fact can be shown explicitly by investigating the dependence of 
ionization and HHG spectra on ellipticity. This is shown in Fig. 2: 
in (a), for a fixed intensity of $0.01 a.u.$ the normalization at the 
end of the pulse is plotted as a function of ellipticity, where the 
results have been scaled such that the normalization for the linear 
case ($\epsilon =0$) equals unity. 
One has to keep in mind that since the intensity is 
kept constant for varying ellipticity, the amplitude of the electric 
field in the initial (linear) polarization direction changes.
The reduced ionization in the initial direction is compensated by the 
possibility of electron ejection in all other directions. The result 
is a net increase of total ionization yield, especially for such high 
intensities that Over the Barrier Ionization (OBI) occurs. This effect 
is even greater if equal electric field amplitudes are compared rather 
than equal intensities. In this case the probability of ionization 
is obviously enhanced, as the electron can escape from the atom by 
tunneling through or over the barrier at any time in the direction 
of the instantaneous electric field. 
    
Clearly, for these parameters we find that as the ellipticity is 
increased the amount of ionization increases by a factor of 
approximately 3. Past experiments \cite{auguste1,augst} 
have found that the threshold for ionization is higher for circular
polarization. However, these experiments have been carried out in the 
tunneling regime where standard tunneling theory \cite{adk}
predicts higher ionization rates for linear polarization. 
The results presented here are not strictly within the tunneling 
regime (due mainly to the relatively high frequency) and represent 
new behaviour. In Fig. 2 (b), we have used the 
same parameters as in Fig. 1 (the spectra are rescaled such that 
the fundamental peak is equal to one in intensity for each ellipticity), 
and now find that the harmonics act as seen in experiments
\cite{elli,liang}: as the ellipticity is increased, the harmonics 
become less intense until no harmonics are discernible from noise. 
Only the first 4 harmonics can be seen for $\epsilon=0.6$, and the peaks 
are found to be rather broad, indicating that they are generated on 
a short timescale. In addition there seems to be a correlation between 
the ionization behaviour and the ellipticity beyond which the 
harmonics are destroyed: the normalization flattens out after 
approximately $\epsilon = 0.7$ which is also the ellipticity at 
which the harmonics disappear.

To highlight the very different behaviour of the ionization for 
various ellipticities we have calculated the normalization at the 
end of the pulse for linear and circular polarization as a function 
of intensity (Fig. 3). At low intensities we see
the expected perturbative multiphoton behaviour, observed in experiment
\cite{liang}, that linear polarization produces a greater amount of 
ionization because of the transition rule restrictions on circular 
polarization transitions. However, as the intensity is increased to 
the critical intensity for over the barrier ionization (around $0.005 
a.u.$ of intensity) we observe that the behaviour changes such 
that circular polarization becomes more efficient in producing 
ionization and indeed for much of the intensity range shown it is 
at least an order of magnitude greater. Note that the normalization 
curve for linear polarization is not smooth: 
this was previously observed by Pindzola and D\"orr \cite{pind} and 
LaGattuta \cite{lag} and assigned to AC Stark shifted resonances and
threshold effects. 
However,we find, as in ref. \cite{lag} that for circular polarization 
these structures do not occur, perhaps because of the more 
restrictive circular dipole selection rules, at least in the perturbative 
regime of intensities.

Finally, we observe a minimum in the normalization 
implying that stabilization is occurring. This is in agreement with 
calculations made by Pont and Gavrila \cite{pont1} for circular 
polarization using {\it time-independent} high frequency Floquet theory. 
The lowest frequency which they have used is $\omega=0.125$ a.u., 
while we have found stabilization, albeit not to the same degree, 
for a frequency 1.5 times smaller. 
Interestingly, for the same parameters linear polarization
shows no signs of stabilization. This may be due to the atom
not surviving the passage through `death valley' \cite{gavr}. 
We are currently investigating adiabatic stabilization 
with much higher frequencies and intensities, the results of which will 
be presented elsewhere.

In summary, we have investigated numerically, 
with two-dimensional ab initio calculations, the dependence of 
ionization and HHG on ellipticity. In particular, we have found 
that the harmonics are destroyed by elliptical light, as 
seen in recent experiments \cite{elli,liang}. 
In a parameter regime where the tunneling picture is not
entirely valid, circular polarization produces a greater amount of 
ionization than linear. We have also found evidence for stabilization 
in circularly polarized light in qualitative agreement with previous
time-independent calculations. The method we propose opens up the 
possibility of studying fully quantum-mechanically wavepacket evolution
in time-dependent ellipticities proposed for the generation of
ultrashort pulses of harmonics \cite{iva}.
As well as this exciting prospect, the unified description 
of the strong-field phenomena, based on the exact wavepacket dynamics 
presented here, will open the way to resolve problems associated with, 
e.g., the accuracy of the `two-step' model for harmonic generation, 
time-dependent laser polarization, control of harmonic generation in 
time, ionization with elliptical light, elliptical stabilization of 
wavepackets, angular distribution of photoelectrons in ATI, and many 
other important questions in strong-field physics that could be 
addressed only by approximate and speculative theoretical models 
in the past.

We would like to thank J.H. Eberly and D.D. Meyerhofer for useful 
discussions. This work has been funded in part by the UK Engineering 
and Physical Sciences Research Council and the European Union.
One of us (D.G.L.) acknowledges support through the EU Training and 
Mobility of Researchers Scheme.

\newpage
\begin{figure}[hbt]
\caption{Snapshots of the probability density taken after 3.5 cycles for
(a) linear polarization and (b) circular polarization
with intensity $3.51 \times 10^{14} Wcm^{-2}$ and wavelength 526 nm.
Distances (X,Y) are shown in atomic units (a.u.).}
\end{figure}

\begin{figure}[hbt]
\caption{(a) Rescaled normalization at the end of the pulse as a function
of ellipticity for a 6 cycle pulse with maximum intensity 
$3.51 \times 10^{14} Wcm^{-2}$ and wavelength 526 nm. (b) Rescaled
HHG spectra as a function of ellipticity for the same laser parameters.}
\end{figure}

\begin{figure}[hbt]
\caption{Normalization as a function of peak laser intensity for a 6 cycle
pulse with wavelength 526 nm, for linear and circular polarization.}
\end{figure}
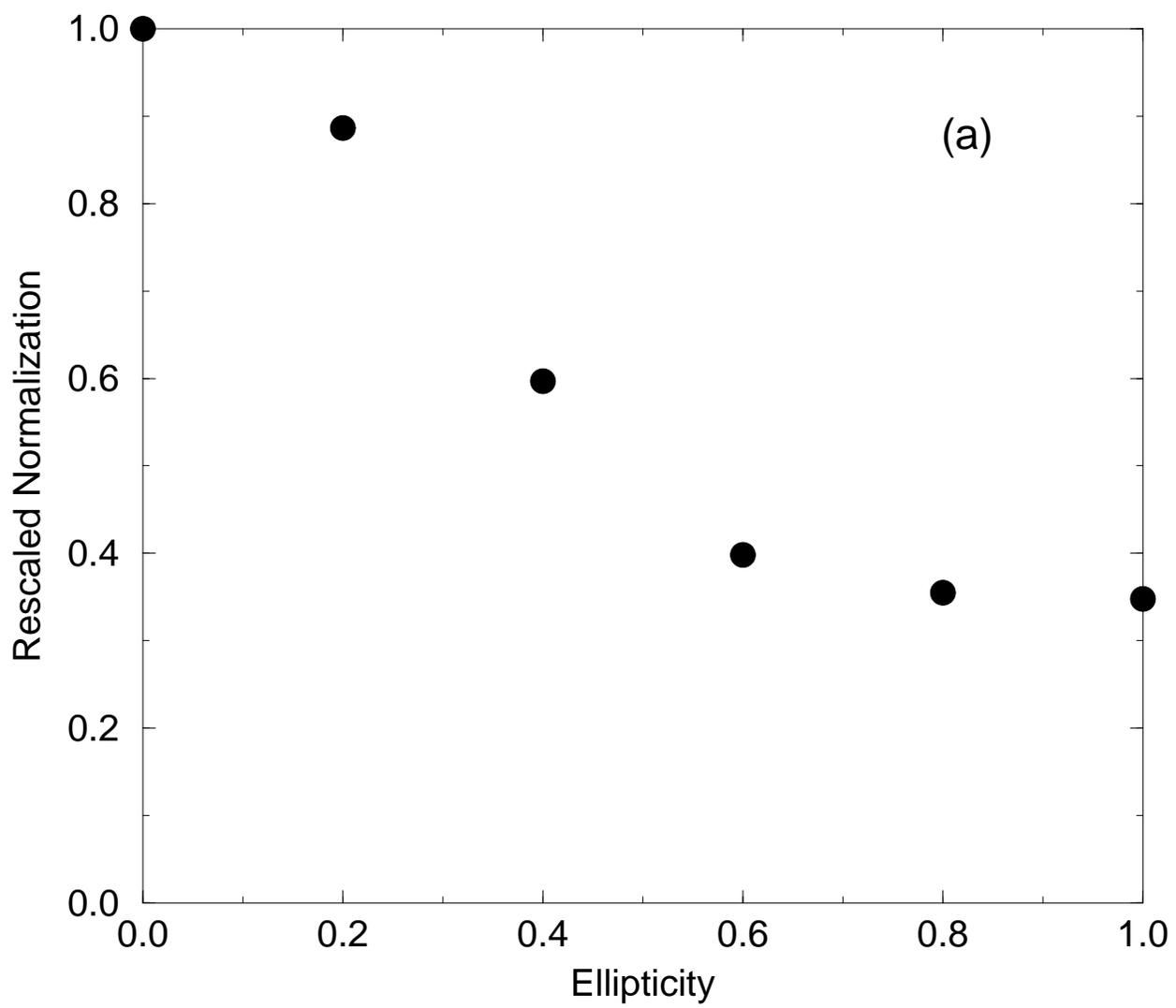
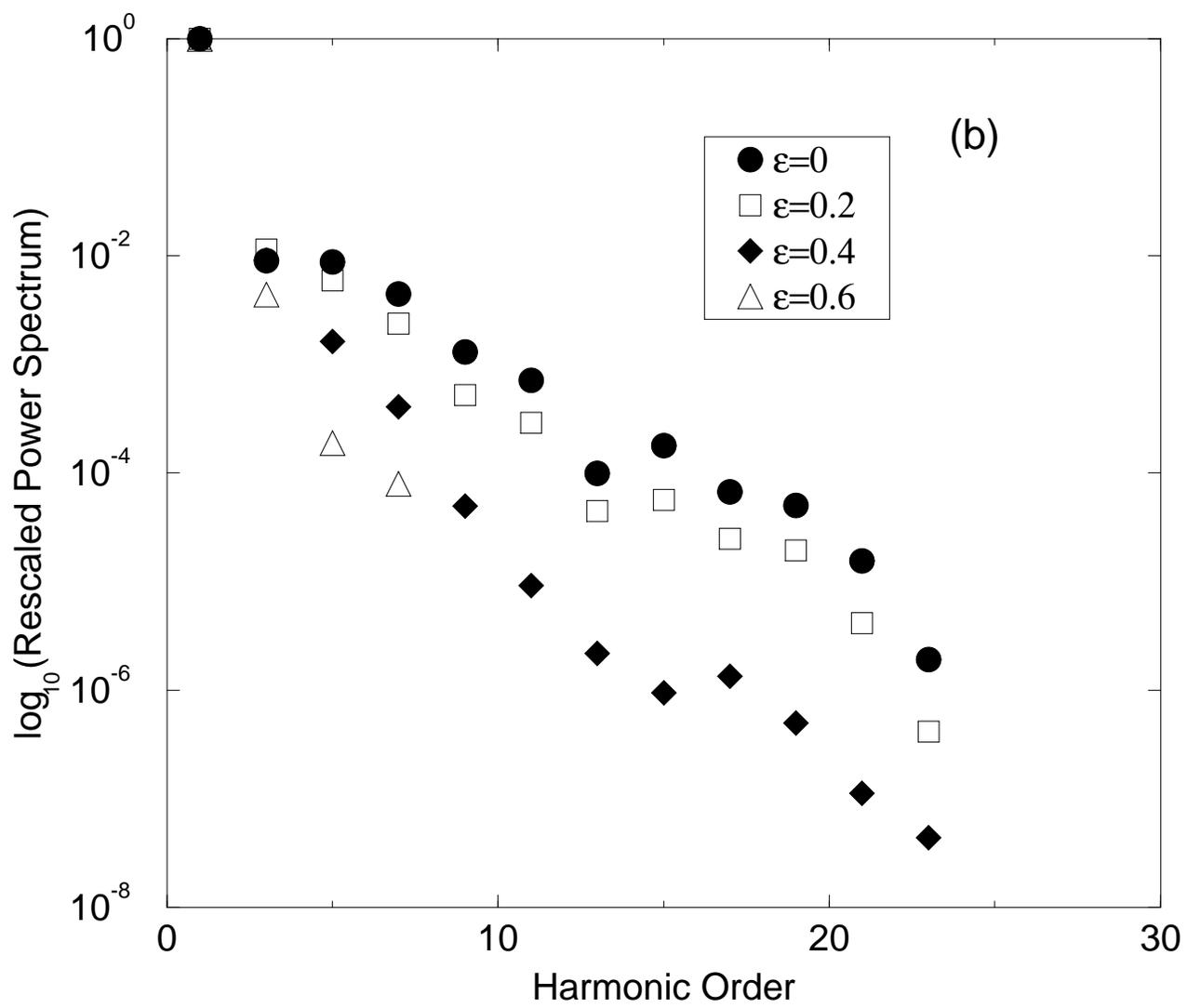
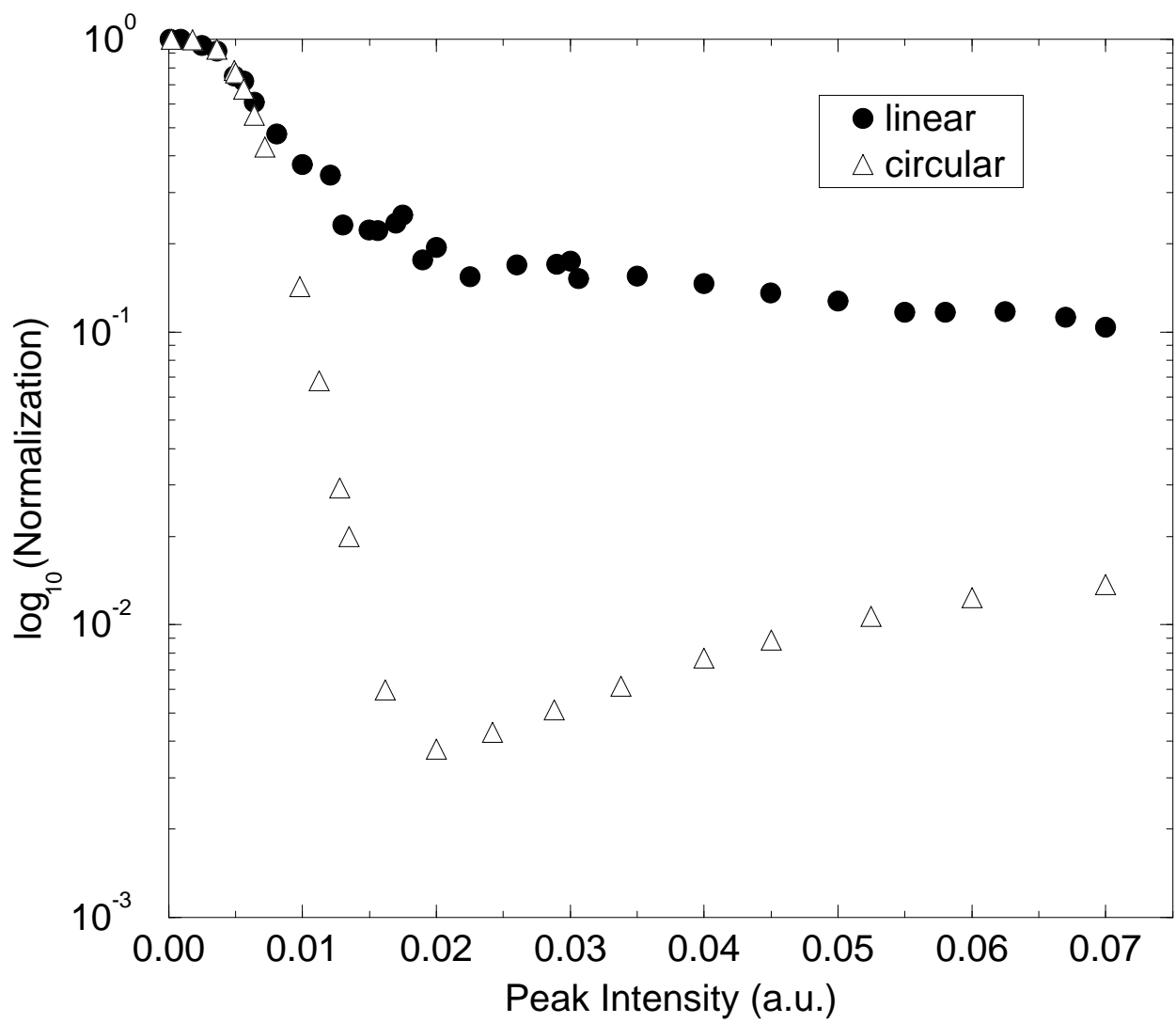
\end{document}